\documentstyle[epsfig,wgsocl]{article}
\newcommand{\simlt}  {\raisebox{-.6ex}{$\stackrel{\textstyle <}{\sim}$}}

\def\be{\begin{equation}}
\def\ee{\end{equation}}
\def\bea{\begin{eqnarray}}
\def\eea{\end{eqnarray}}

\begin{document}

\noindent IDM2000, York, UK. 18-22 Sept.\ 2000. 
\hspace{2.2cm} RAL-TR-2000-047 \\

\title{STATUS OF TRIMAXIMAL NEUTRINO MIXING}

\author{W. G. Scott}

\address{Rutherford-Appleton Laboratory, 
Chilton, Didcot, Oxon. \\ OX11-0QX, UK\\E-mail: w.g.scott@rl.ac.uk}


\maketitle\abstracts{
     Trimaximal mixing is the mixing hypothesis
     with maximal symmetry.
     In trimaximal mixing 
     there remains a still worrying
     conflict between the large values of
     $\Delta m^2$ preferred by the atmospheric fits
     (possibly supported by the early K2K data)
     and reactor limits on $\nu_e$-mixing.
     However, the latest solar results do seem to point to 
     energy-independent (ie.\ `no-scale') solar solutions, 
     like the trimaximal solution.}

\section{Introduction}

At the last IDM meeting (IDM98) I reviewed [1] 
neutrino oscillations,
mentioning trimaximal mixing [2,3] of course,
but giving substantive emphasis
to the so-called bimaximal scheme [4]
which was new at that time.
This year, I sense that bimaximal mixing
is in no particular need of
any `hard-sell' from me,
and I plan therefore
to concentrate on trimaximal mixing,
which arguably merits a little more `air-time' 
than it sometimes gets at this point.

After all, bimaximal mixing
(in its original form) may be seen as just 
the minimal deformation [5] of trimaximal mixing
obtained enforcing a zero
in the top right-hand ($e3$) corner
of the trimaximal mixing matrix
to account for the latest reactor data. 
Evidently (Eq.~1) symmetry between
all three generations
\begin{eqnarray}
\vspace{-0.5cm}
     \matrix{ {\rm trimaximal \hspace{2mm} mixing}  \hspace{3.0cm}
              & {\rm bimaximal \hspace{2mm} mixing} \hspace{1.5cm}  
                                                 } \nonumber \\
     \matrix{  \hspace{1.0cm} \nu_1 \hspace{0.2cm}
               & \hspace{0.2cm} \nu_2 \hspace{0.2cm}
               & \hspace{0.2cm} \nu_3  \hspace{0.2cm} }
\hspace{0.9cm}
     \matrix{  \hspace{2.0cm} \nu_1 \hspace{0.2cm}
               & \hspace{0.2cm} \nu_2 \hspace{0.2cm}
               & \hspace{0.2cm} \nu_3  \hspace{0.4cm} } 
                                \hspace{1.5cm} \nonumber \\
\matrix{ e \hspace{0.0cm} \cr
         \mu \hspace{0.0cm} \cr
         \tau \hspace{0.0cm} }
\left( \matrix{ 1/3  &
                      1/3 &
                              1/3 \cr
                1/3 &
                    1/3 &
                             1/3  \cr
      \hspace{2mm} 1/3 \hspace{2mm} &
         \hspace{2mm}  1/3 \hspace{2mm} &
           \hspace{2mm} 1/3  \hspace{2mm} \cr } \right)
\hspace{2mm} \longleftrightarrow \hspace{2mm}
\matrix{ e \hspace{0.0cm} \cr
         \mu \hspace{0.0cm} \cr
         \tau \hspace{0.0cm} }
\left( \matrix{ 1/2  &
                      1/2 &
                              . \cr
                1/4 &
                    1/4 &
                             1/2  \cr
      \hspace{2mm} 1/4 \hspace{2mm} &
         \hspace{2mm}  1/4 \hspace{2mm} &
           \hspace{2mm} 1/2  \hspace{2mm} \cr } \right) \hspace{1.0cm}
\vspace{1mm}
\end{eqnarray}
is sacrificed in the bimaximal scheme,
but clearly $\nu_1 \leftrightarrow \nu_2$
as well as $\mu \leftrightarrow \tau$
symmetry do survive [5] (note Eq.~1 gives the $|U_{l\nu}|^2$).
It should be mentioned that
the famous Fritzsch-Xing hypothesis [6]
did in fact predict $U_{e3}=0$,
but has otherwise less symmetry than
either the trimaximal or bimaximal schemes.
Altarelli and Feruglio [7] 
usefully generalised
the bimaximal scheme,
retaining the last column of the 
original bimaximal form (Eq.~1 - RHS),
but parametrising the first
two columns in terms of a general mixing angle $\theta$,
to be determined.

In praise of trimaximal mixing, 
the trimaximal mixing matrix (Eq.~1 - LHS)
is clearly especially symmetric, extremal/optimal
and arguably `natural' 
from a number of points of view.
The importance of $Z_3$ symmetry
seems to be generally recognised [8].
In analogy to the Uncertainty Principle
the lepton flavour information is
uniformly spread over the neutrino
mass spectrum and vice versa
(cf.\ also, the famous `hexacode' 
over $F_4$). 
For trimaximal mixing the Jarlskog parameter
$J_{CP}$ [9] takes its extremal value
$J_{CP}=1/(6\sqrt{3})$ such that (vacuum) 
CP and T violating asymmetries are maximised.

\section{Data at the Atmospheric Scale}

The atmospheric neutrino scale
defined by fits to the SUPER-KAMIOKANDE data (Fig.~1) is 
currently $\Delta m^2$ $\simeq$ $3$ $10^{-3}$ eV$^2$
[10].
\begin{figure}[h]
\epsfig{figure=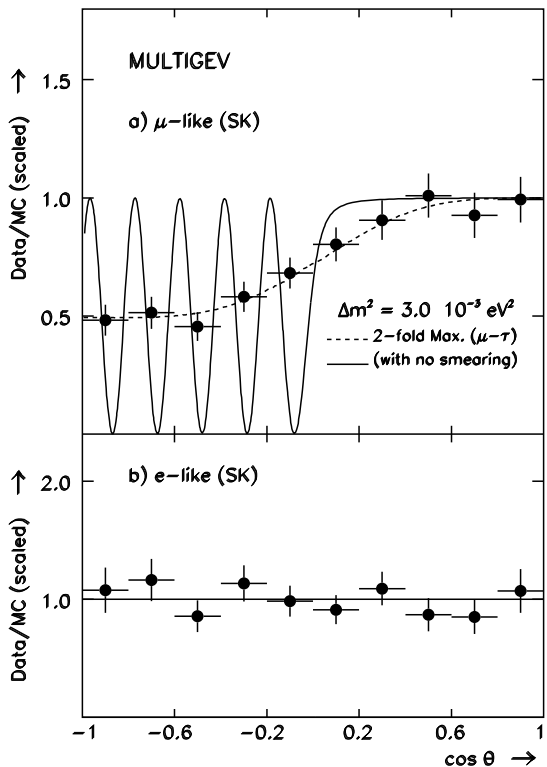,width=100mm,bbllx=0pt,bblly=200pt
,bburx=500pt,bbury=700pt}
\caption{
The multi-GeV zenith angle distributions
for a) $\mu$-like and b) $e$-like events
in SUPER-K.
The solid curve is the full oscillation curve
for bimaximal mixing (Eq.~1 - RHS)
with $\Delta m^2 = 3 \hspace{1mm} 10^{-3}$ eV$^2$
for a representative neutrino energy $E = 3$ GeV.
The dashed curve shows the effect of angular smearing
and averaging over neutrino energies.}
\end{figure}
Upcoming $\mu$-events (Fig.~1a)
are seen to be supressed by a factor 
$P(\mu \rightarrow \mu) \simeq 1/2$. 
Solving the equation $(1-x)^2+x^2=1/2$
yields $x \equiv |U_{\mu 3}|^2 \simeq 1/2$
as in Eq.~1 (RHS).
No deviation 
is seen for $e$-events (Fig~1b), 
but $\phi(\nu_{\mu})/\phi(\nu_e) \simeq 2/1$,
coupled with matter effects
(especially if $\Delta m^2$ $\simlt$ $3$ $10^{-3}$ eV$^2$, see below)
gives low sensitivity to
$\nu_e$-mixing.

Reactor experiments
on the other hand,
specifically CHOOZ [11] and PALO-VERDE[12],
{\em do} rule out large $\nu_e$ mixing
over (almost) all of the $\Delta m^2$-range
currently favoured in the atmospheric 
neutrino experiments.
While the atmospheric experiments claim
$10^{-3}$ eV$^2$  $\simlt$ $\Delta m^2$ $\simlt$ $10^{-2}$ eV$^2$,
at around the $99\%$ confidence level,
the reactor experiments
require $\Delta m^2$ $\simlt$ $10^{-3}$ eV$^2$
unless $U_{e3}$ is small: $|U_{e3}|^2$ $\simlt$ $0.03$.
The near non-overlap of these
two different $\Delta m^2$ ranges
underlies the current popularity of the
(generalised) bimaximal scheme(s) 
discussed above.
\begin{figure}[h]
\epsfig{figure=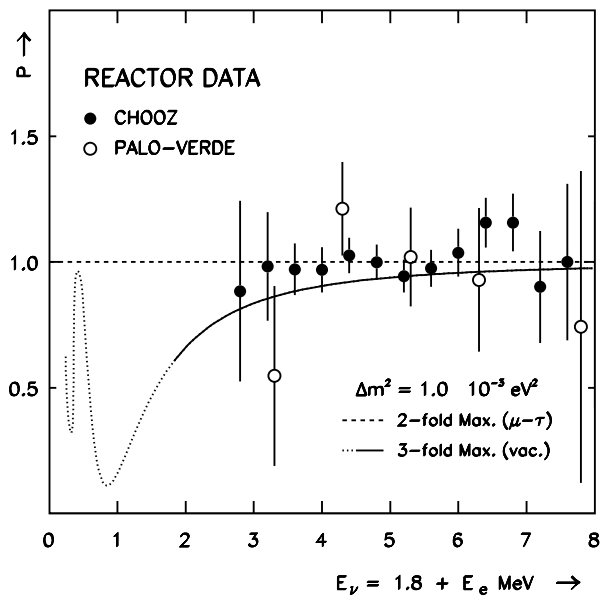,width=100mm,bbllx=0pt,bblly=260pt
,bburx=500pt,bbury=640pt}
\caption{The survival probability 
$P(\bar{\nu}_e \rightarrow \bar{\nu}_e)$
measured in the CHHOZ and PALO-VERDE reactor experiments
(filled and open data points)
compared to the trimaximal mixing prediction
for $\Delta m^2 = 1.0 \hspace{1mm} 10^{-3}$ eV$^2$
(solid curve).}
\end{figure}
Of course trimaximal mixing predicts
$|U_{e3}|^2 = 1/3$ which is large,
so if trimaximal mixing is right
$\Delta m^2$ is at least well determined,
ie.\ $\Delta m^2 \simeq 10^{-3}$ eV$^2$ [3]
with rather little margin for error.
Clearly (Fig.~2) trimaximal mixing
with $\Delta m^2 = 1.0 \hspace{1.5mm} 10^{-3}$ eV$^2$,
would fit the combined CHOOZ and PALO-VERDE data
very nicely, given only a modest re-scaling
of the CHOOZ data by $\sim -8\%$
(the quoted error on the CHOOZ flux
is $\sim \pm 3\%$).

Long-baseline accelerator experiments
like K2K [13]
have well defined $L/E$,
so there is no way
that $\Delta m^2$ is overestimated
by uncertain angular smearing effects
as perhaps in the atmospheric experiments.
Worryingly for trimaximal mixing,
K2K have 27 FC events in fiducial volume with
26.6 events expected
for $\Delta m^2 \simeq 3 \hspace{1mm} 10^{-3}$ eV$^2$,
with very much closer to 40.3 events (the no-oscillation \\
\begin{table}[h]
\vspace{-0.4cm}
\begin{tabular}{lcc}
Conference \hspace{1.6cm}    &  Evts.\ Obs.\ FC IFV \hspace{4mm}
                              & \hspace{2mm} Evts.\ Expd.\ (no osc.)\\
\hline 
QUEBEC        &  ~3        &   12.3   \\
NU2000        &  14        &   16.9   \\
ICHEP2000         &  10        &   11.1   \\
\hline
Total         &  27        &   40.3   \\
\end{tabular}
\caption{K2K fully contained events in fiducial volume, 
seen vs.\ expected (for no oscillation)
(chronologically ordered independent samples 
conveniently separated here by conference.)}
\vspace{-0.3cm}  
\end{table}

\noindent 
expectation)
expected
for $\Delta m^2 \simeq 1.0 \hspace{1.5mm} 10^{-3}$ eV$^2$.
As the K2K experimenters have themselves
pointed out however [13], 
there is something slightly `odd' about the 
distribution of events versus 
chronological expectation (Table~1),
with most of the deficit 
apparently coming from the 1999 running. 

\section{The Solar Data}

The latest SUPER-K data
on the solar supression [14]
extend the electron recoil spectrum
down to $E > 5$ MeV
and start to be convincingly `consistent with flat'
(ie. with energy independence).
Assuming BP98 fluxes [15],
the overall supression $S \simeq 0.47$.
Correcting for the neutral current contribution
(Fig.~3)
we find $P(e \rightarrow e) \simeq 0.38$,
not so very different from the
HOMESTAKE result
$P(e \rightarrow e) \simeq 0.33$.
The solid curve in Fig.~3
is the postdiction of the `optimised'
\begin{eqnarray}
\vspace{-0.5cm}
     \matrix{ {\rm `optimised' \hspace{0mm}}
              & {\rm bimaximal \hspace{2mm} mixing} \hspace{1.5cm}  
                                                 } \nonumber \\
     \matrix{  \hspace{1.0cm} \nu_1 \hspace{0.2cm}
               & \hspace{0.2cm} \nu_2 \hspace{0.2cm}
               & \hspace{0.2cm} \nu_3  \hspace{0.2cm} } 
                                              \hspace{1.5cm} \nonumber \\
( \hspace{1mm} |U_{l\nu} \hspace{1.5mm} |^2) \hspace{1.5mm} = \hspace{1.5mm}
\matrix{ e \hspace{0.0cm} \cr
         \mu \hspace{0.0cm} \cr
         \tau \hspace{0.0cm} }
\left( \matrix{ 2/3  &
                      1/3 &
                              . \cr
                1/6 &
                    1/3 &
                             1/2  \cr
      \hspace{2mm} 1/6 \hspace{2mm} &
         \hspace{2mm}  1/3 \hspace{2mm} &
           \hspace{2mm} 1/2  \hspace{2mm} \cr } \right) \hspace{0.8cm}
\vspace{1mm}
\end{eqnarray} 
bimaximal hypothesis (Eq.~2)
with $\Delta m'^2 = 5.6 \hspace{1.5mm} 10^{-5}$ eV$^2$.
The `optimised' bimaximal form 
is readily obtained from the `generalised' bimaximal scheme [7]
by setting $\sin^2 \theta =1/3$,
and we proposed it [3,16]
only as the best `straw-man' 
rival to trimaximal mixing,
with the possibilty to 
exploit the LA-MSW solution (Fig.~3).
Of course, 
energy {\em in}-dependent solar solutions,
like the trimaximal mixing solution (Fig.~4),
remain a priori much more plausible [16].
It is interesting to see 
that the early SNO data [17] 
support energy independence.



\begin{figure}[h]
\epsfig{figure=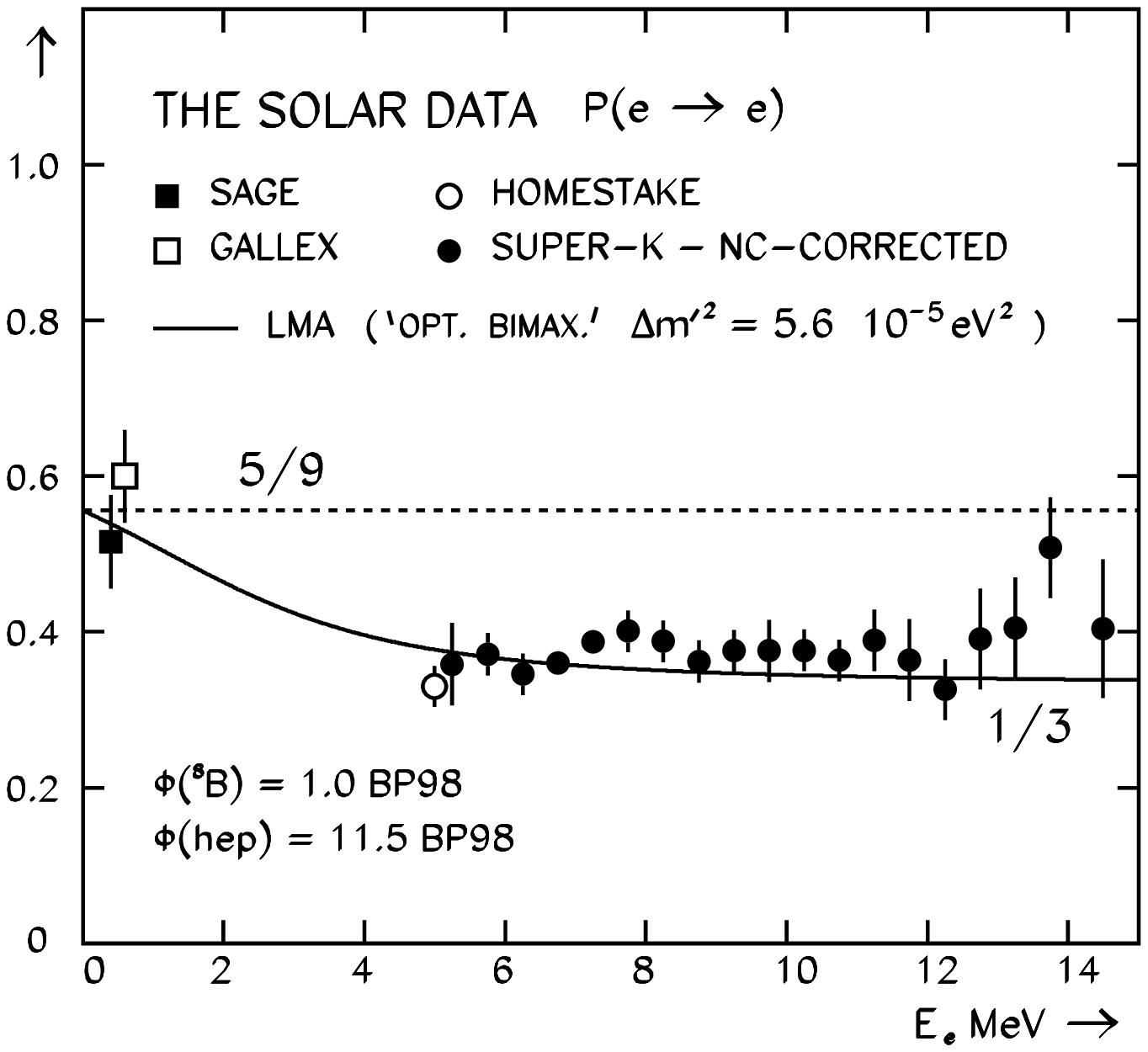,width=80mm,bbllx=-70pt,bblly=260pt
,bburx=380pt,bbury=600pt}
\end{figure}

\noindent \footnotesize 
Figure 3. The SUPER-K solar data [14]
after the NC subtraction
vs.\ recoil energy $E_e$,
assuming the BP98 $^8B$-flux [5]
(with rescaled hep).
SAGE, GALLEX and HOMESTAKE points 
also shown but versus $E_{\nu}$.
The solid curve is `optimised' bimaximal mixing 
(Eq.~2)
with $\Delta m'^2 = 5.6 \times 10^{-5}$ eV$^2$,
giving $P(e \rightarrow e) = 1/3$ 
in the `bathtub' (and $5/9$ otherwise).
\\
 
\begin{figure}[h]
\epsfig{figure=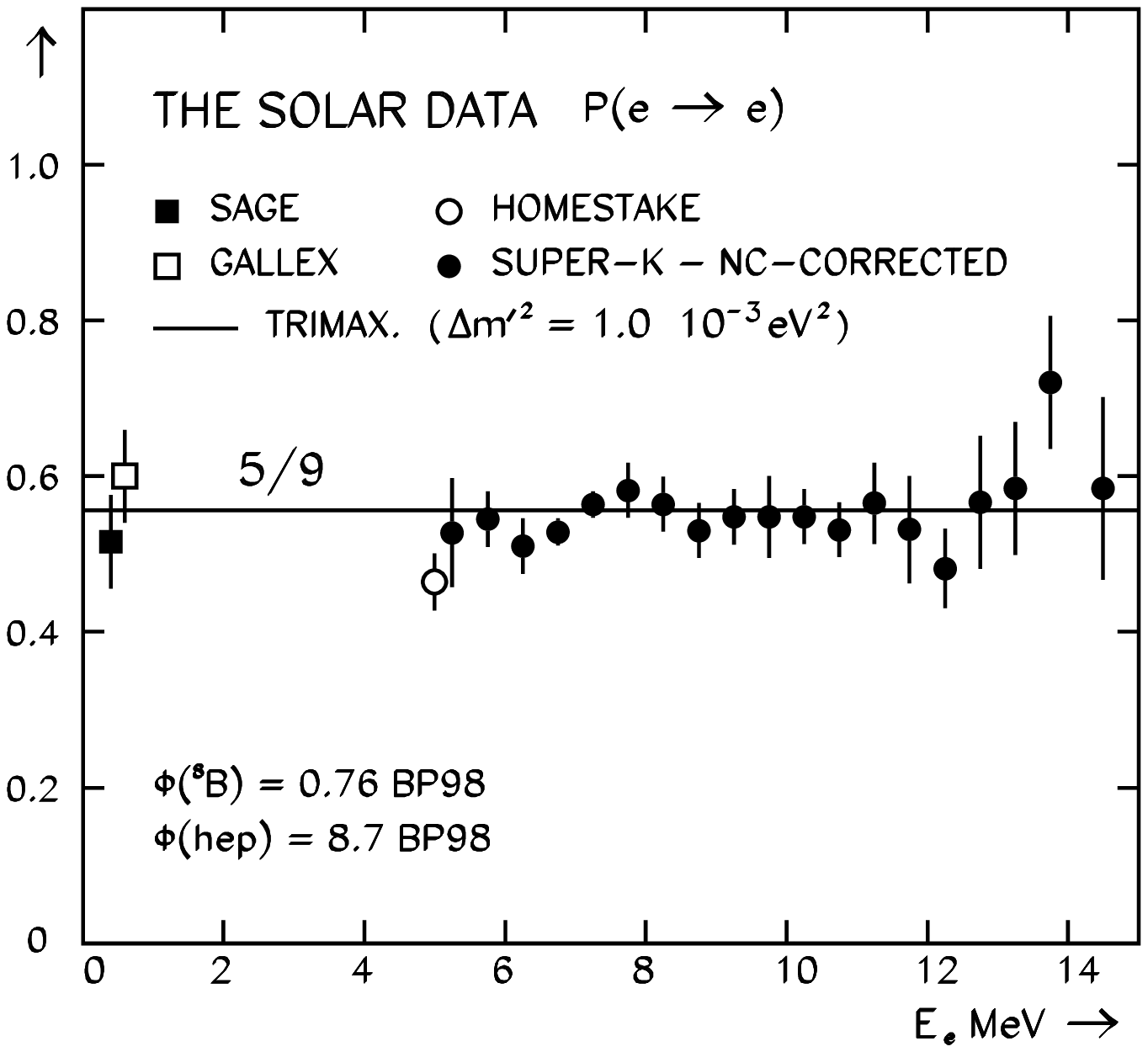,width=80mm,bbllx=-70pt,bblly=260pt
,bburx=380pt,bbury=600pt}
\end{figure}

\noindent
Figure~4. As for Fig.~3, 
except for an arbitrary rescaling of 
BP98 $^8$B-fluc by $-24\%$.
The line is the trimaximal mixing prediction
$P(e \rightarrow e) =5/9$ independent of energy.
Note that Eq.~2
(with or without 
a $\nu_1 \leftrightarrow \nu_2$
column interchange if desired)
likewise gives
$P(e \rightarrow e) =5/9$
independent of energy
outside the `bathtub' region.
Thus Eq.~2 can never be excluded
based on the solar data alone,
underlining again the importance 
of KAMLAND [12], K2K [13] etc.

\newpage

\noindent \normalsize
In SUPER-K, a previous 
$2\sigma$ 
day/night asymmetry
$A = 0.065 \pm 0.031 \pm 0.013$
has fallen with increased statistics 
to $1\sigma$ significance:
$A = 0.034 \pm 0.022 \pm 0.013$ [14].
A day/night effect 
would have been the `smoking-gun'
of the MSW or VO solutions.
Instead, essentially {\em all} 
$10^{-10}$ eV$^2$  $\simlt$ $\Delta m'^2$ $\simlt$ $10^{-3}$ eV$^2$
are now allowed, with (near-)maximal mixing,
{\em except} those explicitly excluded 
(eg.\ $2 \hspace{1.0mm} 10^{-7}$ eV$^2$  
             $\simlt$ $\Delta m'^2$ 
                         $\simlt$ $2 \hspace{1.0mm} 10^{-5}$ eV$^2$)
by the absence of a day/night effect.

\noindent
\section*{Acknowledgment}
It is a pleasure to thank
Paul Harrison and Don Perkins
for continued collaboration/interest
and for helpful comments and suggestions
on this manuscript.


\end{document}